# MINING TARGET-ORIENTED FUZZY CORRELATION RULES TO OPTIMIZE TELECOM SERVICE MANAGEMENT

Hao-En Chueh

Department of Information Management, Yuanpei University, Hsinchu City, Taiwan
hechueh@mail.ypu.edu.tw

*ABSTRACT*

*To optimize telecom service management, it is necessary that information about telecom services is highly related to the most popular telecom service. To this end, we propose an algorithm for mining target-oriented fuzzy correlation rules. In this paper, we show that by using the fuzzy statistics analysis and the data mining technology, the target-oriented fuzzy correlation rules can be obtained from a given database. We conduct an experiment by using a sample database from a telecom service provider in Taiwan. Our work can be used to assist the telecom service provider in providing the appropriate services to the customers for better customer relationship management.*

*KEYWORDS*

*Telecom Services, Target-Oriented Rules, Fuzzy Correlation Rules, Data Mining, Customer Relationship Management*

## 1. INTRODUCTION

Since 1980, most countries have viewed telecommunication development as a vital indicator of social and economic development, causing telecommunications to become one of the most rapidly expanding industries in the world. During this initial period of telecommunications development, the telecom industry in most countries was monopolized by state-owned enterprises. However, to prevent these companies from abusing power and monopolizing markets, international organizations such as General Agreement on Tariffs and Trade (GATT), General Agreement on Trade in Services (GATS), and the World Trade Organization (WTO) held a conference in Uruguay in 1994, and in 1998 this conference reached the Basic Telecommunication Agreement. This agreement established the trend towards privatization, openness, competition, and internationalization.

Regarding the Taiwanese telecom industry, since 1996, when the government officially announced the full opening of the mobile phone communications business to private cable enterprises, many companies have eagerly competed to grab a slice of the emergent telecommunications market. During the initial period of opening the telecom market, users of mobile phone communications services were uncommon; the market demand for this type of emergent communications service was enormous. Consequently, development of the overall mobile phone communications market flourished. However, since then, the consumer market for the telecom industry has changed from vigorously developing to nearly saturated, while remaining fiercely competitive. This has led to the emergence of a serious problem of customer churn. In recent years, as user demand for mobile communications services has changed from rapid growth to increased saturation, many large telecomm service providers have, figuratively speaking, gone from having their own secure customer groups to fierce competition with other providers.

According to statistics, the cost of acquiring one new customer is five to ten times the cost of maintaining one existing customer. To maintain market share and profitability, retaining long-term, loyal customers is more profitable to telecom service providers than acquiring price-conscious, short-term customers. Accordingly, the telecom service providers need to provide the most appropriate telecom services to their customers to maintain customer relationship management. To optimize telecom services management, a useful technique for finding the telecom services, which customers really want, is needed [9, 20].

Of all data mining techniques, mining association rules are often defined by finding the itemsets, which frequently occur, together in a given dataset [1, 2, 13, 15, 21]. The general form of an association rule is $A \Rightarrow B$, where $A$ and $B$ are two itemsets, and $A \cap B = \phi$. A wider application of mining association rules is to identify the customer behavior from various business databases. For example, in the telecommunications industry, mining association rules can help us extract the sets of the telecom services, which are frequently used by the same customers. Clearly, this kind of information is useful for understanding the popular sets of the telecom services that they provide to the customer. However, two services that are frequently used by the same customers, cannot always imply that there exists a strong relationship between them. For example, assuming a telecom service $S$ is popular and used almost by all customers, then the service $S$ is very likely used with other services frequently by some customers, but in fact, the relationship between this service and other services may not be as strong. Thus, mining association rules may generate misleading information for the telecom service provider to provide appropriate services to customers. In this case, mining the correlation rules may be a good way to find the telecom services that customers really want.

Mining correlation rules are procedures for finding the itemsets that are highly related to each other from a given dataset [4, 13, 15]. However, in practical telecom databases, data are not only described by binary values, but also are described by quantitative ones, such as the number of times that a customer used a service in a period time is waiting to be explored. The methods for discovering the correlation rules from these quantitative data are also needed. For avoiding the shape boundary problem, the concept of fuzzy set theory [22] is commonly used to transform these quantitative data into the fuzzy items or the fuzzy attributes, and these are known as fuzzy correlation rules [8, 18, 19].

To discover the services that customers really want, we propose an algorithm for mining the target-oriented fuzzy correlation rules to focus on finding strong correlations between the most popular services and other services according to the numbers of times that customers used these services. Our work can help the telecom service provider to retain the telecom services highly related to the most popular services. In this paper, we show that by using fuzzy statistics analysis and the data mining technology, the target-oriented fuzzy correlation rules can be obtained from a given database. The remainder of this paper is organized as follows. In section 2, the basic concepts involved in mining the fuzzy correlation rules are introduced. In section 3, a proposed algorithm for mining the target-oriented fuzzy correlation rules is presented. In section 4, an experiment is displayed. In section 5, the conclusion is given.

## 2. PRELIMINARIES

Some basic concepts involved in mining the fuzzy association rules and the fuzzy correlation rules are introduced in this section.

### 2.1. Fuzzy Association Rules

Sequential patterns mining which was first introduced by Agrawal and Srikant in the mid 1990s [1], can be described as the task of finding frequently occurring ordered patterns from a given sequence database.

Mining fuzzy association rules is finding the fuzzy itemsets that frequently occur together from a given database [3, 5, 10, 11, 14, 17].

Let $F = \{f_1, f_2, \cdots, f_m\}$ be a set of fuzzy items, $T = \{t_1, t_2, \cdots, t_n\}$; be a set of fuzzy records, and each fuzzy record $t_i$ is represented as a vector with $m$ values, $(f_1(t_i), f_2(t_i), \cdots, f_m(t_i))$, where $f_j(t_i)$ is the degree that fuzzy item $f_j$ appears in record $t_i$, $f_j(t_i) \in [0,1]$. Then, a fuzzy association rule is defined as an implication form, such as $F_A \Rightarrow F_B$, where $F_A, F_B \subset F$ are two fuzzy itemsets, and $\forall_{all\,a} f_a \in F_A \neq \forall_{all\,b} f_b \in F_B$. A fuzzy association rule $F_A \Rightarrow F_B$ holds in $T$ with the fuzzy support, $fsupp(\{F_A, F_B\})$, and with the fuzzy confidence, $fconf(F_A \Rightarrow F_B)$ [12].

The fuzzy support and the fuzzy confidence of the fuzzy association rule, $F_A \Rightarrow F_B$, are defined as

$$fsupp(\{F_A, F_B\}) = \frac{\sum_{i=1}^{n} \min(f_j(t_i) \mid f_j \in \{F_A, F_B\})}{n} \quad (1)$$

$$fconf(F_A \Rightarrow F_B) = \frac{fsupp(\{F_A, F_B\})}{fsupp(\{F_A\})} \quad (2)$$

If $fsupp(\{F_A, F_B\})$ is greater than or equal to the predefined threshold, minimal fuzzy support, and the $fconf(F_A \Rightarrow F_B)$ is also greater than or equal to the predefined threshold, minimal fuzzy confidence, and then the fuzzy association rule, $F_A \Rightarrow F_B$, is considered an interesting fuzzy association rule.

## 2.2. Fuzzy Correlation Rules

While the task of mining fuzzy association rules concerns the co-occurrence of the fuzzy itemsets, the task of mining fuzzy correlation rules focuses on correlation relationships among the fuzzy itemsets. In previous research [8, 18, 19], we have presented an algorithm to discover fuzzy correlation rules by using fuzzy simple correlation analysis [6].

Assume that $F = \{f_1, f_2, \cdots, f_m\}$ is a set of fuzzy items; $T = \{t_1, t_2, \cdots, t_n\}$ is a fuzzy dataset with $n$ records, and each record $t_i$ is represented as a vector with $m$ values, $(f_1(t_i), f_2(t_i), \cdots, f_m(t_i))$, where $f_j(t_i)$ is the degree that fuzzy item $f_j$ occurs in record $t_i$, $f_j(t_i) \in [0,1]$. Then, the form of a fuzzy correlation rule is $F_A \rightarrow F_B$, where $F_A, F_B \subset F$ are two fuzzy itemsets, and $\forall_{all\,a} f_a \in F_A \neq \forall_{all\,b} f_b \in F_B$. A fuzzy correlation rule $F_A \rightarrow F_B$ holds in $T$ with the fuzzy support, $fsupp(\{F_A, F_B\})$, and with the fuzzy confidence, $fconf(F_A \rightarrow F_B)$, and with the fuzzy simple correlation coefficient, $r_{A,B}$.

The fuzzy support and the fuzzy confidence of $F_A \rightarrow F_B$, are the same as the fuzzy association rule. The fuzzy simple correlation coefficient of $F_A \rightarrow F_B$ is defined as

$$r_{A,B} = \frac{s_{A,B}}{s_A \cdot s_B} \quad (3)$$

$$s_{A,B} = \frac{\sum_{i=1}^{n}(\mu_A(t_i) - \overline{\mu_A}) \cdot (\mu_B(t_i) - \overline{\mu_B})}{n-1} \quad (4)$$

$$\mu_A(t_i) = \min(f_j(t_i) \mid f_j \in \{F_A\}) \tag{5}$$

$$\mu_B(t_i) = \min(f_j(t_i) \mid f_j \in \{F_B\}) \tag{6}$$

$$\overline{\mu_A} = fsupp(\{F_A\}) \tag{7}$$

$$\overline{\mu_B} = fsupp(\{F_B\}) \tag{8}$$

$$s_A = \sqrt{S_A^2} \tag{9}$$

$$s_B = \sqrt{S_B^2} \tag{10}$$

$$S_A^2 = \frac{\sum_{i=1}^{n} (\mu_A(t_i) - \overline{\mu_A})^2}{n-1} \tag{11}$$

$$S_B^2 = \frac{\sum_{i=1}^{n} (\mu_B(t_i) - \overline{\mu_B})^2}{n-1} \tag{12}$$

Similarly, if the $fsupp(\{F_A, F_B\})$ of $F_A \to F_B$ is greater than or equal to the minimal fuzzy support; the $fconf(F_A \to F_B)$ of $F_A \to F_B$ is greater than or equal to the minimal fuzzy confidence; the $r_{A,B}$ of $F_A \to F_B$ is also greater than or equal to a predefined threshold, called minimal fuzzy correlation coefficient, and then the fuzzy correlation rule, $F_A \to F_B$, is considered an interesting fuzzy correlation rule.

## 3. TARGET-ORIENTED FUZZY CORRELATION RULES

To discover the strong correlations between a popular telecom service and other telecom services according to the numbers of times that the customer used the service, we propose an algorithm for mining target-oriented fuzzy correlation rules. A target-oriented fuzzy correlation rule is a fuzzy correlation rule with a fuzzy item specified [7, 8]. In this section, we show that the target-oriented fuzzy correlation rules can be obtained from a given database by using fuzzy statistics analysis.

Assume that $F = \{f_1, f_2, \cdots, f_m\}$ be a set of fuzzy items, and $F_t$ is the target fuzzy item of $F$; $T = \{t_1, t_2, \cdots, t_n\}$ be a fuzzy dataset with $n$ records, and each sample record $t_i$ is represented as a vector with $m$ values, $(f_1(t_i), f_2(t_i), \cdots, f_m(t_i))$, where $f_j(t_i)$ is the degree of the fuzzy item $f_j$ belongs to record $t_i$, $f_j(t_i) \in [0, 1]$.

Before staring the algorithm, three user-predefined thresholds, $s_f$, $c_f$, $r_f$, need to be determined, where $s_f$ is the minimal fuzzy support, $c_f$ is the minimal fuzzy confidence, $r_f$ is the minimal fuzzy simple correlation coefficient. The details of the algorithm are described as

Step 1: For each fuzzy item $f_j \in F$, the value of $S(f_j)$ is computed.

$$S(f_j) = \frac{\sum_{i=1}^{n} f_j(t_i)}{n} \tag{13}$$

Step 2: Let $L_1$ be the set of the fuzzy items with value of $S(f_j)$ is greater than or equal to $s_f$. $F_T$ is the fuzzy item with the largest value of $S(f_j)$.

Step 3: Let $L_1' = L_1 - F_T$.

Step 4: Let $C_2$ be the set of the combinations of $L_1'$ joint with $F_T$. $C_2 = \{(F_A, F_T) \mid F_A \in L_1'\}$.

Step 5: For each member of $C_2$, say $(F_A, F_T)$, the value of the fuzzy support, $fsupp(\{F_A, F_T\})$, and the value of the fuzzy simple correlation coefficient, $r_{A,t}$, are computed by formula (1) and by formula (3), respectively.

Step 6: Let $L_2$ is the set of the members of $C_2$ with fuzzy supports are greater than or equal to $s_f$ and fuzzy simple correlation coefficients are greater than or equal to $r_f$.

Step 7: Let $C_2'$ be the set of the combinations of $L_1'$ joint with $L_1'$. $C_2' = \{(F_B, F_C) \mid F_B, F_C \in L_1', F_B \neq F_C\}$.

Step 8: For each member of $C_2'$, say $(F_B, F_C)$, the value of the fuzzy support, $fsupp(\{F_B, F_C\})$, and the value of the fuzzy simple correlation coefficient, $r_{B,C}$, are computed by formula (1) and by formula (3), respectively.

Step 9: Let $L_2'$ is the set of the members of $C_2'$ with fuzzy supports are greater than or equal to $s_f$ and fuzzy simple correlation coefficients are greater than or equal to $r_f$.

Step 10: Generate each $C_k$, $k \geq 3$, by $L_1'$ joint with $L_{k-1}$. Assume that $F_D$ is a member of $L_1'$; $(F_E, F_F)$ is a member of $L_{k-1}$; the size of $F_E$ is $e$. If $(F_E, F_D)$ is a member of $L_{e+1}$ or $L_{e+1}'$, then the combination $(\{F_E, F_D\}, F_F)$ is a member of $C_k$. Similarly, if $(F_F, F_D)$ is a member of $L_{k-e}$ or $L_{k-e}'$, then the combination $(F_E, \{F_F, F_D\})$ is also a member of $C_k$.

Step 11: For each member of $C_k$, say $(F_F, F_G)$, the value of the fuzzy support, $fsupp(\{F_F, F_G\})$, and the value of the fuzzy simple correlation coefficient, $r_{F,G}$, are computed by formula (1) and by formula (3), respectively.

Step 12: Let $L_k$ is the set of the members of $C_k$ with fuzzy supports are greater than or equal to $s_f$ and fuzzy simple correlation coefficients are greater than or equal to $r_f$.

Step 13: Generate each $C_k'$, $k \geq 3$, by $L_{k-1}'$ joint with $L_{k-1}'$. Assume that $(F_P, F_Q)$ and $(F_P, F_R)$ are two members of $L_{k-1}'$; the size of $F_P$ is $p$; the size of the union set, $\{F_Q, F_R\}$, is a $k - p$; and then the combination $(F_P, \{F_Q, F_R\})$ is a member of $C_k'$.

Step 14: For each member of $C_k'$, say $(F_X, F_Y)$, the value of the fuzzy support, $fsupp(\{F_X, F_Y\})$, and the value of the fuzzy simple correlation coefficient, $r_{X,Y}$, are computed respectively by formula (1) and by formula (3).

Step 15: Let $L_k'$ is the set of the members of $C_k'$ with fuzzy supports are greater than or equal to $s_f$ and fuzzy simple correlation coefficients are greater than or equal to $r_f$.

Step 16: Repeat step 10 to step 15, until next $C'_{k+1}$ is an empty set.

Step 17: For each number of $L_k$, $k \geq 2$, say $(F_W, F_Z)$, two candidate fuzzy correlation rules with $F_T$, $F_W \to F_Z$ and $F_Z \to F_W$, can be generated. If the fuzzy confidence of a candidate rule is greater than or equal to $c_f$, then this fuzzy correlation rule is considered an interesting rule.

## 4. EXPERIMENT

A simple dataset sampling came from the records of a telecom company in Taiwan is shown in Table 1. Each entry $t_{ij}$ of Table 1 is the number of times per month that the customer $CID_i$ used the telecom service $S_j$ averaged over eight months.

Before starting the procedure of mining the target-oriented fuzzy correlation rules from this dataset, the thresholds need to be determined in advance. Here, $s_f$ is set to 0.25, $c_f$ is set to 0.80, $r_f$ is set to 0.30. First, we use the formula (14) to determine the degrees of interest to the telecom services for each customer, and transform Table 1 into Table 2. Therefore, each entry $t_{ij}$ of Table 2 is the degree of interest that the customer $CID_i$ used the telecom service $S_j$.

Table 1: A sample telecom dataset.

| $CID_i$ | $S_1$ | $S_2$ | $S_3$ | $S_4$ | $S_5$ |
|---|---|---|---|---|---|
| $CID_1$ | 1.7 | 0.8 | 0.9 | 0.6 | 0.2 |
| $CID_2$ | 1.2 | 8.0 | 1.3 | 0.3 | 1.4 |
| $CID_3$ | 0.8 | 0.8 | 0.4 | 0.1 | 0.6 |
| $CID_4$ | 3.9 | 4.8 | 23.5 | 0.1 | 28.6 |
| $CID_5$ | 3.6 | 0.8 | 51.4 | 6.3 | 0.8 |
| $CID_6$ | 6.2 | 2.1 | 1.8 | 3.8 | 3.0 |
| $CID_7$ | 1.3 | 3.7 | 1.6 | 0.6 | 17.6 |
| $CID_8$ | 33.2 | 20.2 | 14.4 | 2.4 | 4.3 |
| $CID_9$ | 11.7 | 17.7 | 14.8 | 7.5 | 3.8 |
| $CID_{10}$ | 2.8 | 16.8 | 10.7 | 2.4 | 0.2 |
| $CID_{11}$ | 1.3 | 42.3 | 5.4 | 3.5 | 39.1 |
| $CID_{12}$ | 8.8 | 4.4 | 5.4 | 2.4 | 0.3 |
| $CID_{13}$ | 3.6 | 0.5 | 5.1 | 50.7 | 1.5 |
| $CID_{14}$ | 30.7 | 34.7 | 1.6 | 0.8 | 0.2 |
| $CID_{15}$ | 0.4 | 12.4 | 0.9 | 1.2 | 9.1 |
| $CID_{16}$ | 2.5 | 7.4 | 0.8 | 42.9 | 0.4 |

$$\mu_{S_j}(CID_i) = \begin{cases} 1 & , t_{ij} \geq 10 \\ t_{ij}/10 & , 10 > t_{ij} > 0 \\ 0 & , 0 > t_{ij} \end{cases} \quad (14)$$

Table 2: The degrees of interest that the customers used the telecom service.

| $CID_i$ | $S_1$ | $S_2$ | $S_3$ | $S_4$ | $S_5$ |
|---|---|---|---|---|---|
| $CID_1$ | 0.2 | 0.1 | 0.1 | 0.1 | 0.0 |
| $CID_2$ | 0.1 | 0.8 | 0.1 | 0.0 | 0.1 |
| $CID_3$ | 0.1 | 0.1 | 0.0 | 0.0 | 0.1 |
| $CID_4$ | 0.4 | 0.5 | 1.0 | 0.0 | 1.0 |
| $CID_5$ | 0.4 | 0.1 | 1.0 | 0.6 | 0.1 |
| $CID_6$ | 0.6 | 0.2 | 0.2 | 0.4 | 0.3 |
| $CID_7$ | 0.1 | 0.4 | 0.2 | 0.1 | 1.0 |
| $CID_8$ | 1.0 | 1.0 | 1.0 | 0.2 | 0.4 |
| $CID_9$ | 1.0 | 1.0 | 1.0 | 0.8 | 0.4 |
| $CID_{10}$ | 0.3 | 1.0 | 1.0 | 0.2 | 0.0 |
| $CID_{11}$ | 0.1 | 1.0 | 0.5 | 0.4 | 1.0 |
| $CID_{12}$ | 0.9 | 0.4 | 0.5 | 0.2 | 0.0 |
| $CID_{13}$ | 0.4 | 0.1 | 0.5 | 1.0 | 0.2 |
| $CID_{14}$ | 1.0 | 1.0 | 0.2 | 0.1 | 0.0 |
| $CID_{15}$ | 0.2 | 0.1 | 0.1 | 0.1 | 0.0 |
| $CID_{16}$ | 0.1 | 0.8 | 0.1 | 0.0 | 0.1 |

Next, the experimental steps are described as

Step 1: For each telecom service $S_j$, $j = 1 \cdots 5$, the value of $S(S_j)$ is computed and listed in Table 3.

Table 3: $S(S_j)$ of each telecom service $S_j$.

| $S_j$ | $S(S_j)$ |
|---|---|
| $S_1$ | 0.43 |
| $S_2$ | 0.54 |
| $S_3$ | 0.47 |
| $S_4$ | 0.26 |
| $S_5$ | 0.29 |

Step 2: According to Table 3, $L_1 = \{S_1, S_2, S_3, S_4, S_5\}$, and $F_T = S_2$.

Step 3: Let $L_1' = L_1 - F_T$. Thus, $L_1' = \{S_1, S_3, S_4, S_5\}$.

Step 4: Let $C_2$ be the set of the combinations of $L_1'$ joint with $F_t$. Therefore, $C_2 = \{(S_1, S_2), (S_3, S_2), (S_4, S_2), (S_5, S_2)\}$.

Step 5: The fuzzy support and the fuzzy simple correlation coefficient of each member of $C_2$ are computed and listed in Table 4.

80

Table 4: The fuzzy support and the fuzzy simple correlation coefficient
of each member of $C_2$.

| $C_2$ | fsupp | r |
|---|---|---|
| $(S_1, S_2)$ | 0.33 | 0.33 |
| $(S_3, S_2)$ | 0.35 | 0.35 |
| $(S_4, S_2)$ | 0.16 | -0.10 |
| $(S_5, S_2)$ | 0.21 | 0.20 |

Step 6: According to Table 4, $L_2 = \{(S_1, S_2), (S_2, S_3)\}$.

Step 7: Let $C_2'$ be the set of the combinations of $L_1'$ joint with $L_1'$. $C_2' = \{(S_1, S_3), (S_1, S_4), (S_1, S_5), (S_3, S_4), (S_3, S_5), (S_4, S_5)\}$.

Step 8: The fuzzy support and the fuzzy simple correlation coefficient of each member of $C_2'$ are computed and listed in Table 5.

Table 5: The fuzzy support and the fuzzy simple correlation coefficient
of each member of $C_2'$.

| $C'_2$ | fsupp | r |
|---|---|---|
| $(S_1, S_3)$ | 0.31 | 0.45 |
| $(S_1, S_4)$ | 0.19 | 0.31 |
| $(S_1, S_5)$ | 0.14 | -0.15 |
| $(S_3, S_4)$ | 0.22 | 0.42 |
| $(S_3, S_5)$ | 0.20 | 0.26 |
| $(S_4, S_5)$ | 0.11 | 0.03 |

Step 9: According to Table 5, $L_2' = \{(S_1, S_3)\}$.

Step 10: $C_3$ is generated by $L_1'$ joint with $L_2$. Thus, $C_3 = \{(\{S_1, S_3\}, S_2), (S_1, \{S_2, S_3\}), (\{S_2, S_1\}, S_3)\}$.

Step 11: The fuzzy support and the fuzzy simple correlation coefficient of each member of $C_3$ are computed and listed in Table 6.

Table 6: The fuzzy support and the fuzzy simple correlation coefficient
of each member of $C_3$.

| $C_3$ | fsupp | r |
|---|---|---|
| $(\{S_1, S_3\}, S_2)$ | 0.27 | 0.36 |
| $(S_1, \{S_2, S_3\})$ | 0.27 | 0.52 |
| $(\{S_2, S_1\}, S_3)$ | 0.27 | 0.46 |

Step 12: According to Table 6, $L_3 = \{(\{S_1, S_3\}, S_2), (S_1, \{S_2, S_3\}), (\{S_2, S_1\}, S_3)\}$.

Step 13: $C_3'$ is generated by $L_2'$ joint with $L_2'$. $C_3'$ is an empty set, and thus, the step 14, step 15, step 16, are ignored.

Step 17: According to $L_2$ and $L_3$, then candidate fuzzy correlation rules with $S_2$ are generated. The fuzzy confidences of these candidate rules are computed and shown in Table 7.

Table 7: The fuzzy confidences of the candidate target-oriented fuzzy correlation rules generated from $L_2$ and $L_3$.

| Generated Rules | *fconf* |
|---|---|
| $S_1 \rightarrow S_2$ | 0.77 |
| $S_2 \rightarrow S_1$ | 0.61 |
| $S_2 \rightarrow S_3$ | 0.65 |
| $S_3 \rightarrow S_2$ | 0.74 |
| $S_1 \rightarrow \{S_2, S_3\}$ | 0.63 |
| $S_2 \rightarrow \{S_1, S_3\}$ | 0.50 |
| $S_3 \rightarrow \{S_1, S_2\}$ | 0.57 |
| $\{S_2, S_3\} \rightarrow S_1$ | 0.77 |
| $\{S_1, S_3\} \rightarrow S_2$ | 0.87 |
| $\{S_1, S_2\} \rightarrow S_3$ | 0.82 |

According to Table 7, two interesting fuzzy correlation rules with $S_2$ are obtained.

$$\{S_1, S_3\} \rightarrow S_2 \tag{15}$$

$$\{S_1, S_2\} \rightarrow S_3 \tag{16}$$

From rule (15), we can see that there is a strong relationship between telecom services $\{S_1, S_3\}$ and telecom service $S_2$. In addition, rule (16) shows us that there is a strong relationship between telecom services $\{S_1, S_2\}$ and telecom service $S_3$. In the original telecom database, $S_1$ represents the dramatic service; $S_2$ represents the adult service; $S_3$ represents the sport service.

## 5. CONCLUSIONS

To maintain market share and profitability, the telecom service providers need to provide the most appropriate telecom services to their customers to maintain customer relationship management. In this paper, to discover strong correlations between a specific telecom service and other services, we propose an algorithm for mining target-oriented fuzzy correlation rules. Our experimental results can show that there are highly positive relationships among the dramatic service, the adult service and the sport service. Of course, this kind of information is quite useful for the telecom service provider to provide the appropriate services to the customer and remove useless services to reach effective churn management in the telecommunications industry.

**Authors**

Dr. Hao-En Chueh is an Assistant Professor of the Department of Information Management, Yuanpei University, Hsinchu City, Taiwan. He received the Ph.D. in Computer Science and Information Engineering from Tamkang University, Taiwan, in 2007. His research areas include data dining, fuzzy set theory, probability theory, statistics, database system and its applications.

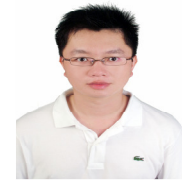